# Game Theoretical Modelling for Dynamic Spectrum Access in TV Whitespace

Ligia Chira Cremene[1,2], D. Dumitrescu[2,3], Réka Nagy[3], Marcel Cremene[1]

[1)] Department of Communications
Adaptive Systems Laboratory
Technical University of Cluj-Napoca, Romania
ligia.cremene@com.utcluj.ro

[2)] Romanian Institute of Science and Technology,
Cluj-Napoca, Romania
www.rist.ro

[3)] Department of Computer Science
Centre for the Study of Complexity
Babes-Bolyai University,
Cluj-Napoca, Romania
ddumitr@cs.ubbcluj.ro

*Abstract*— **In order to assess TV whitespace access scenarios, three oligopoly game models are considered and reformulated in terms of radio access: Cournot, Stackelberg, and Bertrand. Besides revisiting the relevance of Nash and Pareto equilibria, a new equilibrium concept – the joint Nash-Pareto equilibrium is considered. An evolutionary game equilibria detection method is used. The analysis of the simulation results brings relevant insights on the issue of autonomy vs. regulation in emerging cognitive radio environments.**

## I. INTRODUCTION

Current spectrum regimes are based on a highly prescriptive approach, centralized control and decisions. The administrative approach makes it easier for the regulators to ensure avoidance of excessive interference, to tailor appropriate license conditions based on guard bands and maximum power transmission levels [12]. But traditional spectrum planning is valid only for a certain generation of technology. It was proved to be a slow process that cannot keep up with new innovations and technologies. Studies have shown up to 90% of the radio spectrum remains idle in any one geographical location [12], [15]. That is the case with TV whitespaces. Simulations reported by [17] indicate that the DTV whitespaces can provide significantly higher data rates compared to the 2.4 GHz ISM band.

In this paper the problem of harmonized spectrum access in TV whitespaces is addressed. Dynamic spectrum access is a means of managing spectrum enabled through Cognitive Radio [12]. Cognitive radios (CRs) are seen as the solution to the current low usage of the radio spectrum [8], [9], [12]. CRs have the potential to utilize the unused spectrum in an intelligent way while not interfering with other incumbent devices [8], [16]. A cognitive radio has to manage a dynamic interaction profile in a multi-agent environment. This is a suitable task for game theoretic analysis.

Game Theory has been widely used as an analysis tool in economic systems and has recently emerged as an effective framework for the analysis and design of wireless networks. Radio resource allocation and dynamic spectrum access may be described as games between cognitive radios [1], [2], [3], [6]. Widely studied game models are exact potential games [7]. The most frequently used steady-state concept is the Nash Equilibrium (NE) [1], [4], [5]. Yet, there are other equilibria that may be relevant for real access scenarios.

Oligopoly game models are chosen due to the fact that there are few spectrum owners and a large number of users. Three well known oligopoly games are considered: Cournot, Stackelberg, and Bertrand. Besides revisiting the relevance of Nash and Pareto equilibria, a new equilibrium concept – the joint Nash-Pareto equilibrium is considered, in order to more realistically capture radio scene emerging behaviour.

The paper is structured as follows: Section II provides some basic insights to joint game-equilibria generation and detection. The reformulation of the game theoretic models for TV whitespace access is described in Section III. Section IV presents and discusses the numerical results obtained from simulations. The conclusions are presented in Section V.

## II. GAME EQUILIBRIA GENERATION AND DETECTION

A game may be defined as a system $G = ((N, S_i, u_i), i = 1,…, n)$ where:

(i) $N$ represents the set of $n$ players, $N = \{1,…, n\}$.

(ii) for each player $i \in N$, $S_i$ represents the set of actions $S_i = \{s_{i1}, s_{i2}, …, s_{im}\}$; $S = S_1 \times S_2 \times … \times S_N$ is the set of all possible game situations;

(iii) for each player $i \in N$, $u_i : S \to R$ represents the payoff function.

A strategy profile (strategy or action vector) is a vector $s = (s_1,…, s_n)$, where $s_i \in S_i$ is a strategy (or action) of player $i$. By $(s_i, s_{-i}^*)$ we denote the strategy profile obtained from $s^*$ by replacing the strategy of player $i$ with $s_i$, i.e. $(s_i, s_{-i}^*) = (s_1^*, s_2^*,…, s_{i-1}^*, s_i, s_{i+1}^*,…, s_1^*)$.

Informally, a strategy profile is an NE if no player can improve her payoff by unilateral deviation.

Considering two strategy profiles $x$ and $y$ from $S$, the strategy profile $x$ is said to Pareto dominate the strategy profile $y$ (and we write $x < P\ y$) if the payoff of each player using strategy $x$ is greater or equal to the payoff associated to strategy $y$ and at least one payoff is strictly greater. The set of all non-dominated strategies (Pareto frontier) represents the set of Pareto equilibria of the game [5].

In an $n$-player game consider that each player $i$ acts based on a certain type of rationality $r_i$, $i = 1,…, n$. We may consider a two-player game where $r_1$ = Nash and $r_2$ = Pareto. The first player is biased towards the Nash equilibrium and the other one is Pareto-biased. Thus, a new type of equilibrium, called the joint Nash-Pareto equilibrium, may be considered [4]. The considered generalization involves heterogeneous players that are biased towards different equilibrium types or may act based on different types of rationality [4].

Let us consider an $n$ player game where each player may be either Nash or Pareto-biased. We denote by $I_N$ the set of Nash biased players ($N$-players) and by $I_P$ the set of Pareto biased players ($P$-players). Therefore we have

$$I_N = \{i \in \{1,...,n\}: r_i = Nash\},$$
$$I_P = \{j \in \{1,...,n\}: r_j = Pareto\}$$

An operator for measuring the relative efficiency of profile strategies has been introduced [5].

$$E: S \times S \to N,$$

defined as

$$E(y, x) = card\{i \in I_N : u_i(x_i, y_{-i}) \geq u_i(y), x_i \neq y_i\}$$
$$+ card\{j \in I_P : u_j(y) < u_j(x), x \neq y\} .$$

$E(x, y)$ measures the relative efficiency of the strategy profile $x$ with respect to the strategy profile $y$. The relative efficiency enables us to define a generative relation for the joint Nash-Pareto (NP) equilibrium.

Consider a relation $<NP$ defined as $y <NP\ x$ if and only if

$$E(y, x) < E(x, y).$$

The relation $<NP$ is considered as the generative relation of the joint Nash-Pareto equilibrium.

An evolutionary technique for equilibria detection, based on appropriate generative relations [4] that allow the comparison of strategies, is considered.

Numerical experiments aim the detection of pure equilibria or a combination of equilibria paralleling cognitive radios interaction. An adaptation of the popular NSGA2 [11] has been considered.

## III. OLIGOPOLY MODELS FOR TV WHITESPACE ACCESS

In order to assess whitespace access scenarios, three oligopoly game models are considered and reformulated in terms of radio access: Cournot, Stackelberg, and Bertrand [5]. An oligopoly is a market form in which a market or industry is dominated by a small number of sellers (oligopolists).

In the Cournot economic competition model players simultaneously choose quantities. In the Stackelberg model they move sequentially (some have priority) and in the Bertrand's oligopoly they simultaneously choose prices [5]. The commodity of this oligopoly market is the frequency spectrum. In order to illustrate spectrum access situations, scenarios with two radios trying to access the same whitespace (set of channels) are considered. Radio strategies and payoffs are represented two-dimensionally.

### A. Cournot spectrum access modelling

We consider a general spectrum access scenario that can be modelled as a reformulation of the Cournot oligopoly game [1], [5]. Suppose there are $n$ radios attempting to access the same whitespace at the same time. Each radio $i$ may decide the number $c_i \in [0, \infty)$ of simultaneous channels to access. The question is how many simultaneous channels should each radio access in order to maximize its operation efficiency in a harmonized manner?

Based on the above scenario, a Cournot game can be reformulated according to Table 1:

TABLE I.  COURNOT GAME REFORMULATION

| Players | cognitive radios attempting to access a certain whitespace $W$; |
|---|---|
| Actions | the strategy of each player $i$ is the number $c_i$ of simultaneous accessed channels; A strategy profile is a vector $c = (c_1,...,c_n)$. |
| Payoffs | the difference between a function of goodput $P(c)c_i$ and the cost of accessing $c_i$ simultaneous channels $Kc_i$. |

We consider a linear inverse demand function in which the number of non-interfered symbols $P(c)$ is determined from the total number $c_i$ of accessed channels (occupied bandwidth).

The demand function can be defined as:

$$P(C) = \begin{cases} W - C, if C < W, \\ 0, otherwise, \end{cases}$$

where $W > 0$ is the parameter of the inverse demand function and $C = \sum_{i=1}^{n} c_i$, is the aggregate number of accessed channels.

The goodput for radio $i$ is $P(c)c_i$. Radio $i$'s cost for supporting $c_i$ simultaneous channels is $C_i(c_i)$.

The payoff of radio $i$ may then be written as:

$$u_i(c) = P(c)c_i - C_i(c_i) .$$

In general, $P$ decreases with the total number of implemented channels and $C_i$ increases with $c_i$ (more bandwidth implies more processing resources and more power consumption) [1]. If these effects are approximated as linear functions, the payoff function can be rewritten as

$$u_i(c) = \left( W - \sum_{k=1}^{n} c_k \right) c_i - Kc_i ,$$

where
$W$ is the whitespace (set of available channels),
$K$ is the cost of accessing one channel.

The Nash equilibrium is considered as the solution of this game and can be calculated as follows:

$$c_i^* = (W - K)/(n+1), \forall i \in N.$$

*B. Stackelberg spectrum access modelling*

A spectrum access situation where we have incumbent monopoly (licensed users) and new entrants (unlicensed), may well be modelled using a Stackelberg game model where the users move sequentially (one has priority over the other).

The Stackelberg reformulation of the game is captured by Table II:

TABLE II. STACKELBERG GAME REFORMULATION

| Players | the cognitive radios – licensed and unlicensed (primary and secondary) users attempting to access a set of channels; |
|---|---|
| Actions | the strategy of each player $i$ is the number $c_i$ of accessed channels; |
| Payoffs | the difference between a function of goodput and the cost of accessing $c_i$ channels. |

Using the same notations as for Cournot modelling, the payoff function of user $i$, in case of Stackelberg model, can be defined as

$$u_i(c_1, c_2) = c_i P_d(c_1 + c_2) - C_i(c_i), \text{ for } i = 1,2.$$

considering $c_2 = b_2(c_1)$ as the output of the secondary user for primary user's output $c_1$ [6]. We consider a constant unit cost and a linear inverse demand function $P_d(c)$ with the same definition as for the Cournot model.

The outcome of the equilibrium [5] is that radio 1 accesses $c_1^* = \frac{1}{2}(W - K)$ simultaneous channels and radio 2 accesses $c_2^* = b_2(c_1^*) = \frac{1}{4}(W - K)$ simultaneous channels.

*C. Bertrand spectrum access modelling*

In the Bertrand competition, producers compete by varying the product price and thus adjusting the demand. A constant unit cost and linear demand function are assumed. The Bertrand competition for whitespace access may be reformulated as captured by Table III.

We consider $n$ cognitive radios competing for access to $c_i$ channels, in a given whitespace $W$. The objective of each radio is to activate a subset of channels in order to satisfy its current demand level (e.g. target throughput). Using the same notations as for Cournot and Stackelberg models, the payoff function of radio $i$ can be defined as

$$u_i(p_1, p_2) = (p_i - K)(W - p_i), p_i < p_j$$
$$= 1/2(p_i - K)(W - p_i), p_i = p_j$$
$$= 0, p_i > p_j.$$

TABLE III. BERTRAND GAME REFORMULATION

| Players | the cognitive radios attempting to access the whitespace |
|---|---|
| Actions | the strategy of each player $i$ is a target number $p_i(c)$ of non-interfered symbols; |
| Payoffs | the difference between a function of goodput and the cost of accessing $c_i$ simultaneous channels. |

IV. NUMERICAL EXPERIMENTS

The results represent a sub-set of more extensive simulations. For equilibria detection the evolutionary technique from [4] is considered. A population of 100 strategies has been evolved using a rank based fitness assignment technique. In all experiments the process converges in less than 20 generations. Our tests show that the evolutionary method for equilibrium detection is scalable with respect to the number of available channels [13].

The simulation parameters for all the three models – Cournot, Stackelberg, and Bertrand – are $W = 10$ (available channels) and $K = 1$ (cost of accessing one channel).

*A. Cournot modelling – numerical experiments*

Model evaluation results are presented for the Cournot competition with two radios trying to access the same whitespace at the same time. The emerging behaviour of the radio scene is captured by the detected equilibria (Fig. 1): Nash, Pareto, Nash-Pareto, and Pareto-Nash. The four types of equilibria are obtained in separate runs.

The NE corresponds to the scenario where each of the two CRs activates 3 channels (from 10 available). The Pareto equilibrium describes a more unbalanced situation where the number of active channels for each CR lies in the range [0, 4.5].

Although each CR tries to maximize its utility, none of them can access more than half of the available channels. The NE indicates the maximum number of channels a Nash-biased CR may access without decreasing its payoff (Fig. 2).

In some cases, the Nash-Pareto strategy enables the CR to access more channels than for the NE strategy. In the performed experiments the PN equilibrium is symmetric to the NP equilibrium with respect to the first bisecting line. It is interesting to notice that none of the NP strategies actually reach NE.

Fig. 2 illustrates the payoffs of the two players $u_1(c_1, c_2)$ and $u_2(c_1, c_2)$, for the Cournot modeling.

The payoffs of P-players are in the range [0, 20] and their sum is always larger than the NE payoff (9,9). The Nash-player's payoff is smaller in a Nash-Pareto situation than in a case where all the players are Nash-biased.

Even if the NP strategies allow the CRs to access more channels, the payoffs are smaller than for the Pareto strategies. This is due to interference increasing with the number of accessed channels.

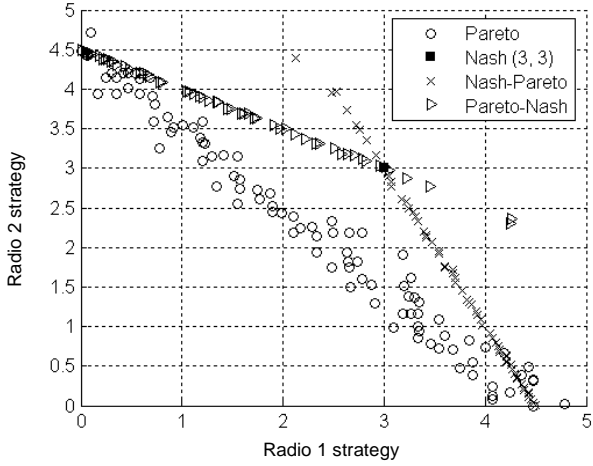

Figure 1.  Cournot modelling – two radios (W = 10, K = 1). Evolutionary detected equilibria: Nash (3,3), Pareto, Nash-Pareto, and Pareto-Nash

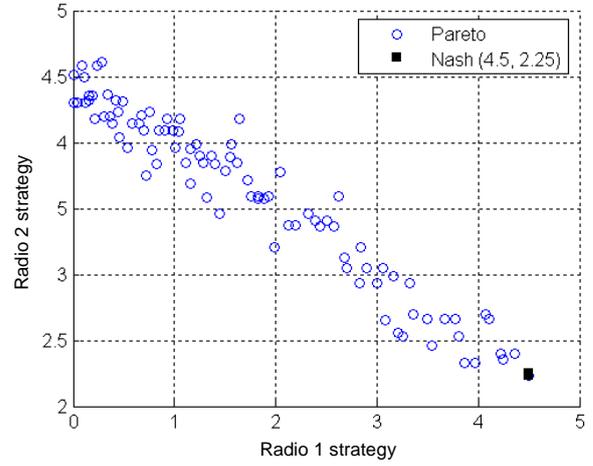

Figure 3.  Stackelberg modelling – two radios (W=10, K=1). Evolutionary detected equilibria: Nash (4.25, 2.25), Pareto, Nash-Pareto, and Pareto-Nash.

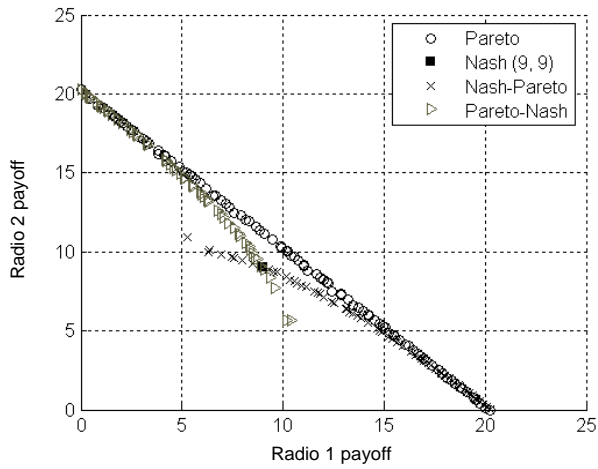

Figure 2.  Cournot modelling – two radios (W = 10, K = 1). Payoffs of the evolutionary detected equilibria: Nash (9, 9), Pareto, NP, and PN.

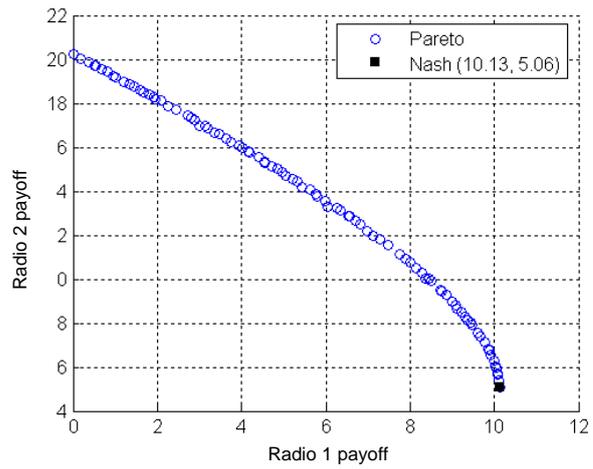

Figure 4.  Stackelberg modelling – two radios (W=10, K=1). Payoffs of the evolutionary detected equilibria: Nash (10.13, 5.06), Pareto, NP=Nash, and PN=Pareto.

*B. Stackelberg modelling – numerical experiments*

The evolutionary detected Nash, Pareto, NP, and PN equilibria are captured in Fig. 3. We may notice that any strategy from the Pareto front is also an NP strategy. If the primary user plays Nash, then the secondary user may maximize its payoff by choosing any strategy. If the secondary user plays Nash then the maximum payoff of the primary user is NE (10.13, 5.06) (Fig.4).

Even if the secondary user can access less channels than in the Cournot case – $c_2$ = 2.25 (Fig.3) which is less than three channels, NE = (3,3) (Fig.1), its maximum payoff remains unaffected, 20 (Fig.4). Instead, the primary user's maximum payoff is half, 10, even if it accesses more channels ($c_1$ = 4.5). For the Stackelberg formulation of the game, the NE payoff of the secondary user (Fig. 4) is less then in the Cournot case (5 instead of 9). For the primary user the NE payoff is slightly increased (10.13 instead of 9) compared to Cournot.

This situation is relevant for interference control in dynamic spectrum access scenarios between incumbents and new entrants. The analysis shows that payoffs are maximized for all users if the incumbents are Nash oriented and the new entrants are Pareto driven.

*C. Bertrand modelling – numerical experiments*

We think the Bertrand oligopoly is suitable for modelling crowded spectrum access scenarios and the reformulation is as follows. The Bertrand strategy is the price. The equivalent of the price $P(c)$ in this game reformulation is the target number of non-interfered symbols of each radio. The lower this target is the higher the chances are for the radio to access one or several channels. On the other hand, as the number $P(c)$ of non-interfered symbols per channel decreases, the need for channels (the demand) increases. Thus, a radio willing to maximize its goodput will attempt to occupy as many low-rate channels as possible. Fig. 5 and Fig. 6 qualitatively illustrate the winning situations for two radios trying to access a limited bandwidth $W$. The NE in this case means zero payoff for each radio while the Pareto strategy ensures the maximum possible payoff for one radio at a time. This indicates that, for a high interference scene, some sort of scheduling or sequential access scheme is required.

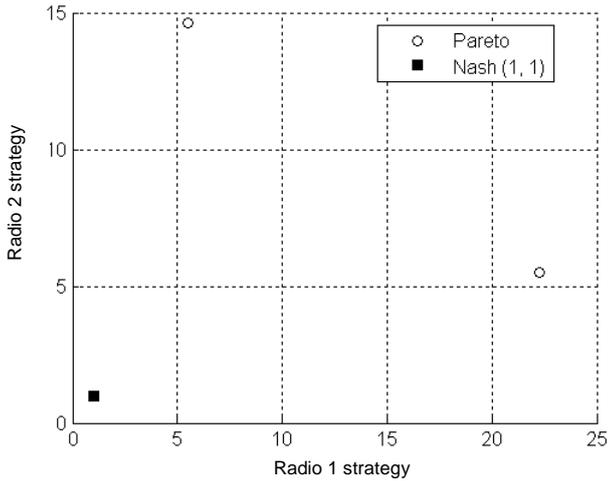

Figure 5. Bertrand modelling – two radios (W=10, K=1). Evolutionary detected equilibria: Nash (1, 1), Pareto.

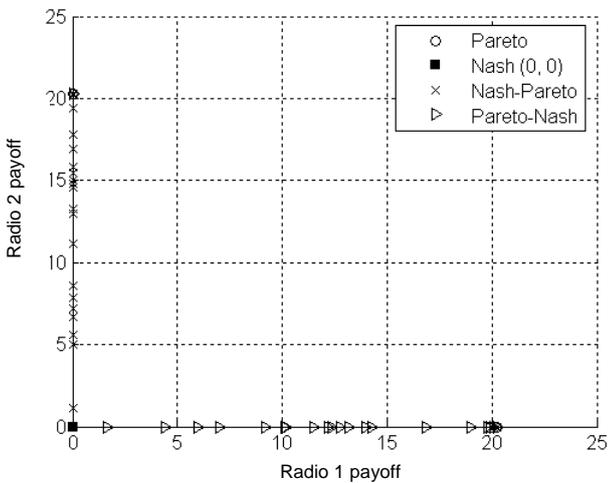

Figure 6. Bertrand modelling – two radios (W=10, K=1). Payoffs of the evolutionary detected equilibria: Nash (0, 0), Pareto, NP, and PN.

## V. CONCLUSIONS

In order to investigate the relevance of certain game equilibrium concepts for the problem of dynamic spectrum access in cognitive radio environments three oligopoly game models were considered. Besides the standard Nash equilibrium new equilibrium concepts were investigated: the Pareto equilibrium and the joint Nash-Pareto equilibrium. The Nash equilibrium indicates the maximum number of channels a Nash-biased CR may access without decreasing its payoff. The analysis of the Stackelberg modelling shows that payoffs are maximized for all users if the incumbents are Nash oriented and the new entrants are Pareto driven. The Bertrand model proves valuable in estimating the chances in a win-lose situation, in a very crowded spectrum. The observations may be especially relevant for designing new rules of behaviour for heterogeneous radio environments.

ACKNOWLEDGMENT

This paper was supported by CNCSIS –UEFISCDI, Romania, PD, project number 637/2010. This work was also supported by the project "Develop and support multidisciplinary postdoctoral programs in primordial technical areas of national strategy of the research - development - innovation" 4D-POSTDOC, contract nr. POSDRU/89/1.5/S/52603, project co-funded from European Social Fund through Sectorial Operational Program Human Resources 2007-2013. The second author acknowledges the support of a grant from the John Templeton Foundation. The opinions expressed in this publication are those of the authors and do not necessarily reflect the views of the John Templeton Foundation.